\documentclass{iopart}
\usepackage{graphicx,bm,iopams}

\newcommand{\ba}{\begin{eqnarray}} \newcommand{\ea}{\end{eqnarray}}
\newcommand{\be}{\begin{equation}} \newcommand{\ee}{\end{equation}}
\newcommand{\nn}{\nonumber}

\renewcommand{\d}{\textrm{d}} 
\newcommand{\diag}{\textrm{diag}} 
 \renewcommand{\i}{\textrm{i}}
 
\newcommand{\s}{\sigma} \renewcommand{\e}{\epsilon}
\newcommand{\eps}{\epsilon} 
\renewcommand{\l}{\lambda}

\newcommand{\fr}[2]{{\textstyle{\frac{#1}{#2}}}}

\begin{document}

\title{The eccentric universe}

\author{Arjun Berera*\ , Roman V. Buniy\dag\ and\\ Thomas
W. Kephart\ddag}

\address{*\ School of Physics, University of Edinburgh, Edinburgh, EH9
3JZ, United Kingdom}

\address{\dag\ Institute of Theoretical Science, University of Oregon,
Eugene, OR 97403, USA}

\address{\ddag\ Department of Physics and Astronomy, Vanderbilt
University, Nashville, TN 37235, USA}

\address{E-mail: ab@ph.ed.ac.uk, roman@uoregon.edu and
thomas.w.kephart@vanderbilt.edu}




\begin{abstract}
For a universe containing cosmological constant together with uniform
arrangements of magnetic fields, strings, or domain walls, exact
solutions to the Einstein equations are shown to lead to a universe
with ellipsoidal expansion. The magnetic field case is the easiest to
motivate and has the highest possibility of finding application in
observational cosmology.
\end{abstract}



\section{Introduction}

The first year WMAP results
\cite{Bennett:2003bz,Spergel:2003cb,Hinshaw:2003ex} contain
interesting large-scale features which warrant further attention
\cite{Tegmark:2003ve,deOliveira-Costa:2003pu}. First, there is the
suppression of power at large angular scales ($\theta \gtrsim
60^{\circ}$), which is reflected most distinctly in the reduction of
the quadrupole $C_2$. This effect was also seen in the COBE
results~\cite{Smoot:1992td,Kogut:1996us}.  However, WMAP was able to
make more precise measurements of $C_2$. Next, the quadrupole $C_2$
and octupole $C_3$ are found to be aligned.  In particular, the
$\ell=2$ and $3$ powers are found to be concentrated in a plane
inclined about $30^{\circ}$ to the Galactic plane.  In a coordinate
system in which the equator is in this plane, the $\ell= 2$ and $3$
powers are primarily in the $m=\pm \ell$ modes. The axis of this
system defines a unique direction and supports an idea of power in one
direction being suppressed compared to the power in the orthogonal
plane. These effects seem to suggest one (longitudinal) direction may
have expanded differently from the other two (transverse) directions,
where the transverse directions describe the equatorial plane
mentioned above.

After the COBE experiment, Monte Carlo studies were
used~\cite{Berera:1997wz} to cast doubt on quadrupole suppression and
analyses of the WMAP data
\cite{Bennett:2003bz,Spergel:2003cb,Hinshaw:2003ex,Tegmark:2003ve,deOliveira-Costa:2003pu}
have arrived at similar conclusions. Nevertheless, interesting
physical effects are not ruled out and a realistic physical model that
could explain some or all effects in the WMAP data would be of
interest.

It would be most satisfying to explain global anisotropy in the
universe by a simple modification of the conventional
Friedman-Robertson-Walker (FRW) model. As a first step in this
direction, we will consider an energy-momentum tensor which is
spatially non-spherical or spontaneously becomes non-spherical. Such
situations could occur when defects or magnetic fields are
present. Magnetic fields \cite{Kronberg:1993vk} and cosmic defects
\cite{Hindmarsh:1994re} can arise in various ways. Moreover, large
scale magnetic fields exist in the universe, perhaps up to
cosmological scales \cite{Kronberg:1993vk,Wick:2000yc}. These
considerations motivate us to examine the effect magnetic fields and
defects have on the expansion of the universe~\cite{RZNB}.

To begin to understand the form, significance and implications of an
asymmetric universe, we here modify the standard spherically symmetric
FRW cosmology to a form with only planar symmetry
\cite{Taub:1950ez}. Our choice of energy-momentum will result in
non-spherical expansion from a spherical symmetric initial
configuration: an initial co-moving sphere will evolve into an
ellipsoid that can be either prolate or oblate depending on the choice
of matter content. We then explore the effects of this type of
asymmetric expansion on modes of a multipole expansion. For the sake
of clarity, we first give some general properties of cosmologies with
planar symmetry (The universe looks the same from all points but they
all have a preferred direction.) and then investigate one case in
detail. Our exemplar will be a universe filled with uniform magnetic
fields and cosmological constant.  This is perhaps the most easily
motivated, analytically solvable case to consider and it will give us
a context in which to couch the discussion of other examples with
planar symmetry and cases where planar symmetry is broken.

To set the stage, consider an early epoch in the universe at the onset
of cosmic inflation, where strong magnetic fields have been produced
in a phase transition
\cite{Savvidy:1977as,Matinian:1976mp,Vachaspati:nm,Enqvist:1994rm,Berera:1998hv}.
The magnitude of the magnetic field and vacuum energy ($\Lambda$)
densities initially are about the same, but eventually $\Lambda$
dominates. It was estimated~\cite{Vachaspati:nm} that the initial
magnetic field energy produced in the electroweak phase transition was
within an order of magnitude of the critical density. Other phase
transitions may have even higher initial field values
\cite{Enqvist:1994rm,Berera:1998hv}. Hence, it is not unphysical to
consider a universe with B-fields and $\Lambda$ of comparable
magnitudes. If the B-fields are aligned in domains, then some degree
of inflation is sufficient to push all but one domain outside the
horizon. (Below we also discuss cases where there are still a few
domains within the horizon.)

\section{Planar symmetric cosmology}

To make the simplest directionally anisotropic universe, we modify
spherical symmetry of FRW space-time into planar
symmetry. (Cylindrical symmetry is, of course, not appropriate since
it introduces a preferred location of the axis of symmetry.) The most
general form of planar-symmetric metric (up to a conformal
transformation) is~\cite{Taub:1950ez}
$(g_{\mu\nu})=\diag{(1,-e^{2a},-e^{2a},-e^{2b})},\label{metric}$ where
$a$ and $b$ are functions of $t$ and $z$; and the $xy$-plane is the
plane of symmetry. We also impose translational symmetry along the
$z$-axis; the functions $a$ and $b$ now depend only on $t$.  For the
above metric, the nonzero-components of the Ricci tensor then are \ba
&&{R^0}_0=-(2\ddot{a}+\ddot{b}+2\dot{a}^2+\dot{b}^2),\label{Ricci00}\nn\\
&&{R^1}_1={R^2}_2=-(\ddot{a}+2\dot{a}^2+\dot{a}\dot{b}),\label{Ricci1122}\nn\\
&&{R^3}_3=-(\ddot{b}+\dot{b}^2+2\dot{a}\dot{b}).\label{Ricci33}\nn\ea

To support the symmetry of space-time, the energy momentum tensor for
the matter has to have the same symmetry: \ba ({T^\mu}_\nu)=(8\pi
G)^{-1} \diag(\xi,\eta,\eta,\zeta).\label{T}\ea Here the energy
density $\xi$, transverse $\eta$ and longitudinal $\zeta$ tension
densities are functions of time only.  Examples of plane-symmetric
spaces include space uniformly filled with either magnetic fields,
static aligned strings, or static stacked walls; i.e., the defects are
at rest with respect to the cosmic background frame. Of course, we can
add to these any spherically-symmetric contributions: vacuum energy,
matter, radiation (see Table~{\ref{table}}). The entries in
Table~{\ref{table}} are scaled according to Eq.~(\ref{T}); for
instance, $\l=8\pi G\Lambda$.

\begin{table}[htb]
\caption{\label{table}Various contributions to the components of the
energy-momentum tensor in Eq.~(\ref{T}).}
\begin{center}
\begin{tabular}{lp{1cm}cp{0.5cm}cp{0.5cm}c}\hline\hline 
{\rule[-2mm]{0mm}{6mm} }& & $\xi$ & & $\eta$ & & $\zeta$\\ \hline
{\rule[4mm]{0mm}{0mm}vacuum energy} & & $\l$ & & $\l$ & & $\l$\\
matter & & $\rho$ & & $0$ & & $0$\\ radiation & & $\rho$ & &
$-\fr{1}{3}\rho$ & & $-\fr{1}{3}\rho$\\ magnetic field & & $\e$ & &
$-\e$ & & $\e$\\ strings & & $\e$ & & $0$ & & $\e$\\
{\rule[-3mm]{0mm}{2mm}walls} & & $\e$ & & $\e$ & & $0$\\ \hline\hline
\end{tabular}
\end{center}
\end{table}

The Einstein equations corresponding to Eq.~(\ref{T}) are
\ba&&\dot{a}^2+2\dot{a}\dot{b}=\xi,\label{A1}\\
&&\ddot{a}+\ddot{b}+\dot{a}^2+\dot{a}\dot{b}+\dot{b}^2=\eta,\label{A2}\\
&&2\ddot{a}+3\dot{a}^2=\zeta.\label{A3}\ea In addition, we need the
equation expressing covariant conservation of the energy-momentum [a
direct consequence of Eqs.~(\ref{A1})--(\ref{A3})]: \ba
\dot{\xi}+2\dot{a}(\xi-\eta)+\dot{b}(\xi-\zeta)=0.
\label{A4}\ea
 
Before considering specific models for the energy-momentum, we first
establish several general features of an anisotropic universe
described by Eqs.~(\ref{A1})--(\ref{A4}). We assume that before
anisotropic effects became important, the universe was
isotropic~\cite{zero}, $a_\i=b_\i$, and was expanding isotropically,
$\dot{a}_\i=\dot{b}_\i>0$. Without loss of generality we set $a_\i=0$,
which is equivalent to setting the initial scale factor to
unity. During the initial phase of anisotropic expansion when
non-uniform contributions to the energy-momentum are more significant,
different tension densities in the longitudinal and transverse
directions cause the universe to expand non-spherically. At a later
time, when all contributions except for the vacuum energy fade away,
longitudinal and transverse expansion rates become equal and the
expansion again proceeds isotropically. Thus, an initial co-moving
sphere evolves into an ellipsoid; whether it is oblate ($a>b$) or
prolate ($a<b$) depends on which tension dominated during the initial
phase of deformation.

For all known forms of classical matter, the components of the
energy-momentum tensor satisfy the dominant energy
condition~\cite{energy-momentum}; in our case it says $\xi\ge 0$,
$\xi\ge\eta$, and $\xi\ge\zeta$ (see Table~\ref{table} for
examples). These conditions lead to various bounds and asymptotics of
which we list here a few important examples.

First, one can show that for the initial conditions given above, the
energy density never increases. Also, the space always expands
transversally, and unless the magnetic field contribution dominates,
the space expands longitudinally. Next, the transverse expansion rate
has its maximum at $t=t_\i$ and the same is true for the longitudinal
expansion rate unless the wall contribution dominates.

Second, the eccentricity~\cite{ecc} is bounded. Indeed,
Eqs.~(\ref{A1}) and (\ref{A3}) lead to $\dot{a}-\dot{b}
=(\zeta-\xi-2\ddot{a})/2\dot{a}$, which upon using the energy
condition $\xi\ge\zeta$ gives the bound
$e^{a-b}\le\dot{a}_{\i}/\dot{a}$. Solving Eq.~(\ref{A3})
asymptotically for large $t$, we find
$\dot{a}\sim(\zeta/3)^{\frac{1}{2}}$, which leads to the asymptotic
bound \ba e^{a-b} \lesssim
(\xi_\i/\zeta)^{\frac{1}{2}}.\label{ab-ineq1a}\ea Note that in an
expanding universe $\xi_{\i}\ge\xi\ge\zeta$ and so (\ref{ab-ineq1a})
does not forbid $a>b$. A sub-case with vacuum energy and any
combination of magnetic fields and strings aligned in the same
direction has $\xi=\zeta$, and similarly we find \be
e^{a-b}\sim(\xi_\i/\zeta) ^{\frac{1}{2}}.\label{A:ab-asymptotic}\ee

Third, let us expand Eqs.~(\ref{A4}) for large $t$ to find
\ba(\delta\xi)\,\dot{}+n(\fr{1}{3}\l)^{\frac{1}{2}}\delta\xi=0;\ea
here $n=3-2\delta\eta/\delta\xi-\delta\zeta/\delta\xi$. For
cosmological constant plus magnetic field, strings, or walls, $n=4,2$
or $1$, respectively. Thus the energy density approaches its
asymptotic value $\l$ according to $\delta\xi\propto e^{-t/t_\xi}$
with characteristic time $t_\xi=n^{-1}(3/\l)^{\frac{1}{2}}$.

We consider here three exactly solvable cases: cosmological constant
plus either uniform magnetic field, homogeneously aligned strings, or
homogeneously stacked walls.  We concentrate on the case of magnetic
fields, which is the easiest to motivate.

\section{Magnetic field}

In the case of magnetic fields, conservation of energy-momentum,
Eq.~(\ref{A4}), has a simple form: $\dot{\e}+4\dot{a}\e=0$. Solving
this equation for $\dot{a}$ and substituting the result into
Eq.~(\ref{A3}), we arrive at the equation
$\e\ddot{\e}-\fr{11}{8}\dot{\e}^2+2\e^2(\l+\e)=0$ whose general
solution is given implicitly by \ba t-t_\i=\fr{1}{4}
\int^{\e_\i}_{\e}d\e\left(\fr{1}{3}\l\e^2+\fr{4}{3}\e_\i^{\frac{1}{4}}
\e^{\frac{11}{4}}-\e^3\right)^{-\frac{1}{2}}.\label{B:integral}\ea
Once the function $\e$ is found from Eq.~(\ref{B:integral}), the
functions $a$ and $b$ are found from Eqs.~(\ref{A1}) and (\ref{A4}) as
follows: 
\begin{eqnarray}
&&a=\fr{1}{4}\ln{(\e_\i/\e)},\label{B:aa}\\ &&
b=\fr{1}{2}\ln\frac{\l+4\eps_\i(\eps/\eps_\i)^\frac{3}{4}-3\eps}
{\l+\eps_\i}-\fr{1}{4}\ln{(\eps/\eps_\i)}.\label{B:bb}
\end{eqnarray}

To find the behavior of $a$ and $b$ for large $t$ we need to find
asymptotics for the integral in Eq.~(\ref{B:integral}). The integral
diverges for small $\e$ (which corresponds to large $t$), and so we
extract the divergent part first; this results in
$t-t_\i\approx\fr{1}{4}(3/\l)^{\frac{1}{2}}\ln{(\e_\i/\e)}-\tau$,
where \ba\tau=\fr{1}{4}\int_{0}^{\e_\i}
d\e\left[\left(\fr{1}{3}\l\e^2\right)^{-\frac{1}{2}}
-\left(\fr{1}{3}\l\e^2+\fr{4}{3}\e_\i^{\frac{1}{4}}
\e^{\frac{11}{4}}-\e^3\right)^{-\frac{1}{2}}\right].\label{B:tau}\nn\ea
Large $t$ asymptotics give \ba&&\e\sim \e_\i\,
e^{-4(\l/3)^{\frac{1}{2}}(t-t_\i+\tau)},\label{B:e}\\ &&a\sim
(\l/3)^{\frac{1}{2}}(t-t_\i+\tau),\label{B:a}\\ &&b\sim
(\l/3)^{\frac{1}{2}}(t-t_\i+\tau)-\fr{1}{2}
\ln{\left(1+\e_\i/\l\right)}.\label{B:b}\ea Eqs.~(\ref{B:a}) and
(\ref{B:b}) lead to an oblate expansion with
$e^{a-b}\sim(1+\e_\i/\l)^{\frac{1}{2}}$, which agrees with the general
result found in Eq.~(\ref{A:ab-asymptotic}).

We can include matter and still find exact solutions. For example, for
a universe filled with cosmological constant, magnetic fields and dust
with density $\rho$, we find
\begin{eqnarray}
\rho=\rho_\i(\eps/\eps_\i)^\frac{3}{4}\left[1+F(\eps/\eps_\i)\right]^{-1}
\left[\frac{\l
+(\rho_\i+4\eps_\i)(\eps/\eps_\i)^\frac{3}{4}-3\eps}{\l+\rho_\i+\eps_\i}
\right]^{-\frac{1}{2}},
\label{B:rho0}
\end{eqnarray}
where
\begin{eqnarray}
F(\eps/\eps_\i)
=&&\fr{3}{8}(\rho_\i/\eps_\i)[1+(\l+\rho_\i)/\eps_\i]^{\frac{1}{2}}\nn\\
&&\int_{\eps/\eps_\i}^{1}\d x\,x^{-\frac{1}{4}}
\left[\l/\eps_\i+(4+\rho_\i/\eps_\i)x^{\frac{3}{4}}-3x\right]^{-\frac{3}{2}}.
\label{B:F}
\end{eqnarray}
Equation~(\ref{B:aa}) is unchanged but Eq.~(\ref{B:bb}) becomes
\begin{eqnarray}
b&&=\fr{1}{2}\ln\frac{\l+(\rho_\i+4\eps_\i)(\eps/\eps_\i)^\frac{3}{4}-3\eps}
{\l+\rho_\i+\eps_\i}\nn\\
&&-\fr{1}{4}\ln{(\eps/\eps_\i)}+\fr{1}{2}\ln\left[1+F(\eps/\eps_\i)\right],
\label{B:b0}
\end{eqnarray}
and time dependence of the above functions $\rho(\eps)$, $a(\eps)$ and
$b(\eps)$ can be deduced from the function $\eps(t)$, which is now
given implicitly by
\begin{eqnarray}
t-t_\i=\fr{1}{4} \int^{\eps_\i}_{\eps}\d\eps\left[
\fr{1}{3}\l\eps^2+\fr{1}{3}\eps_\i^{-\frac{3}{4}} (4\eps_\i+\rho_\i)
\eps^{\frac{11}{4}}-\eps^3\right]^{-\frac{1}{2}}.\label{B:integral0}
\end{eqnarray}

\begin{figure}
\begin{center}
\includegraphics[width=8cm]{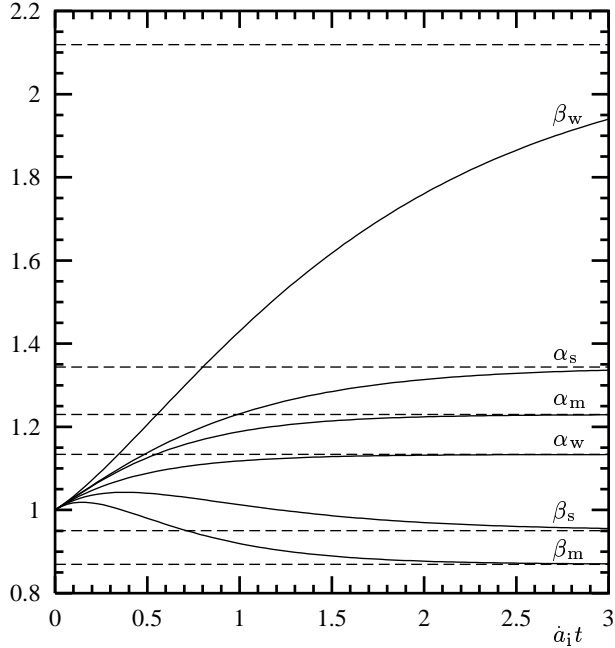}
\caption{The scale factors $\alpha=e^{a-c}$ and $\beta=e^{b-c}$, where
$c=(\fr{1}{3}\l)^{\frac{1}{2}}(t-t_\i)$ corresponds to the model with
only cosmological constant. For the cases of magnetic field, strings,
and walls, the same initial conditions are used. Solid lines are exact
solutions and dashed lines are asymptotics for large
$t$.}\label{fig-ab}
\end{center}
\end{figure}

\begin{figure}
\begin{center}
\includegraphics[width=8cm]{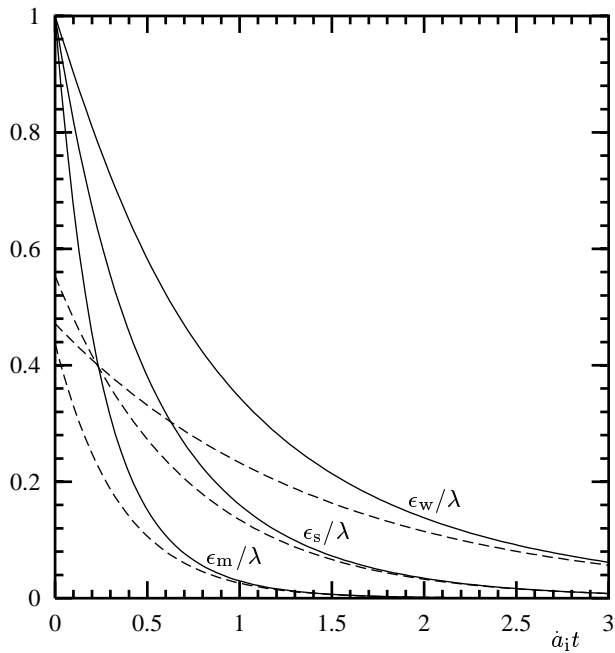}
\caption{The ratio $\e/\l$ for the same three cases and initial
conditions as in Fig.~\ref{fig-ab}.}\label{fig-e}
\end{center}
\end{figure}

\section{Strings and Walls}

For strings, solutions to the Einstein equations can be obtained
similarly to the magnetic field case. Here we give only large $t$
behavior of these solutions: \ba&&\e\sim \e_\i\,
e^{-2(\l/3)^{\frac{1}{2}}(t-t_\i+\tau)},\label{S:e}\nn\\ &&a\sim
(\l/3)^{\frac{1}{2}}(t-t_\i+\tau),\label{S:a}\nn\\ &&b\sim
(\l/3)^{\frac{1}{2}}(t-t_\i+\tau)-\fr{1}{2}
\ln{\left(1+\e_\i/\l\right)},\label{S:b}\nn\ea where
\ba\tau=\fr{1}{2}\int_{0}^{\e_\i}
d\e\left[\left(\fr{1}{3}\l\e^2\right)^{-\frac{1}{2}}
-\left(\fr{1}{3}\l\e^2-\fr{2}{3}\e_\i^{-\frac{1}{2}}
\e^{\frac{7}{2}}+\e^3\right)^{-\frac{1}{2}}\right].\label{S:tau}\nn\ea
As in the case of magnetic field,
$e^{a-b}\sim(1+\e_\i/\l)^{\frac{1}{2}}$, in agreement with
Eq.~(\ref{A:ab-asymptotic}).

Solving the case of walls proceeds in a different fashion. Asymptotic
forms of the corresponding solutions read \ba&&\e\sim
\e_\i\,e^{-(\l/3)^{\frac{1}{2}}
(t-t_\i)}\left[\frac{(1-\s)^{\frac{1}{3}}}{1+\s} +\frac{6\s
F}{(1-\s)^2}\right]^{-1},\label{W:e}\nn\\ &&a\sim
(\l/3)^{\frac{1}{2}}(t-t_\i)-\fr{2}{3}\ln{(1-\s)},\label{W:a}\nn\\
&&b\sim(\l/3)^{\frac{1}{2}}(t-t_\i)
+\ln\left[\frac{(1-\s)^{\frac{1}{3}}}{1+\s} +\frac{6\s
F}{(1-\s)^2}\right],\label{W:b}\nn\ea where \ba\s
=\frac{\left(1+\e_\i/\l\right)^{\frac{1}{2}}-1}
{\left(1+\e_\i/\l\right)^{\frac{1}{2}}+1}\label{W:sigma}\nn\ea and \ba
F=\int_1^\infty d x\,\frac{(x^3-\s)^{\frac{4}{3}}}{(x^3+\s)^2}.\nn\ea
One can derive the bound
$F\ge\fr{1}{3}(1-\s)^{\frac{4}{3}}(1+\s)^{-1}$, which leads to a
prolate expansion with $e^{a-b}\lesssim 1$.

In Fig.~\ref{fig-ab} we compare the scale factors for the cases of
magnetic field, strings, and walls. Since in all three cases the
functions $a$ and $b$ asymptotically approach the corresponding
solution for the model with only cosmological constant,
$c=(\fr{1}{3}\l)^{\frac{1}{2}}(t-t_\i)$, this term is conveniently
factored out of the scale factors. Fig.~\ref{fig-e} similarly compares
the corresponding energy densities. In both figures we used the same
initial conditions for the three cases. These conditions are fully
specified by the value $\e_\i/\l$; we chose $\e_\i=\l$ which leads to
$\dot{a}_\i=(\fr{2}{3}\l)^{\frac{1}{2}}$.

\section{Conclusions}

Since the time of COBE, various speculations have been put forward to
explain the origin of the suppressed quadrupole. One idea has been to
explore topological features with specific emphasis on the ``small
universe'' picture with a toroidal $T^3$ topology~\cite{ellis}.
Another idea has been to study the possibility of a small inflation,
which somehow gets tuned to about the scale of the present-day Hubble
radius~\cite{Berera:1997wz,Cline:2003ve,Bastero-Gil:2003bv}.  However,
very little has been suggested for physical models that could explain
the new features suggested by the WMAP data. An obvious modification
of the $T^3$-model would be to make one dimension smaller than the
other two and get a $T_a^2 \times T_b^1$ asymmetric toroidal topology
\cite{deOliveira-Costa:2003pu}. Other topologies could also
work. There are many choices \cite{Gibbons:1996aq,Ratcliffe:rb} and
the results could be profound.

In this paper we have not departed radically from conventional
cosmology, but rather considered a simple modification of the FRW
model. We determined how magnetic fields, strings, and walls affect
the expansion rate of the universe. Before solving the Einstein
equations we found the asymptotic values and bounds for the ratio of
longitudinal and transverse sizes for several interesting cases. Next,
we presented exact analytical solutions for a universe filled with
magnetic field and gave asymptotic results for the case of defects.
Finally, we mentioned particle physics motivated models that could
yield the sorts of magnetic fields and/or defects of importance for
asymmetric expansion. We have found that a universe filled with
aligned B-fields and $\Lambda$ expands non-spherically: a co-moving
sphere at $t_\i$ evolves into an ellipsoid. The eccentricity is
calculable and fixed by the initial conditions. Aside from being
potentially applicable to explaining some features in the WMAP data,
the models and the results presented in this paper differ from those
in the already mentioned Ref.~\cite{RZNB}. The first of the papers in
Ref.~\cite{RZNB} examined the situation where there is a uniform
magnetic field and no cosmological constant. By adding cosmological
constant, we have found the exact solution for the model which is
physically relevant. The second paper in Ref.~\cite{RZNB} treats only
small anisotropies and determines only asymptotic large time
behavior. Our calculations apply to anisotropy of arbitrary strength
and we have obtained the exact solutions to the Einstein equations
which are valid at all times.

The effects of non-spherical expansion and small inflation are
potentially measurable depending on the degree of eccentricity and
expansion attained and the quality of data. In realistic particle
physics motivated models, where the B-field in question is associated
with particular gauge fields and the cosmological constant $\Lambda$
arises from some specific inflaton model, additional predictions may
emerge. For example, a strongly aligned B-field at the beginning of
inflation may affect density perturbations in a measurable
way~\cite{Tsagas:1999ft}, perhaps through polarization effects, and a
small inflation that follows may mean these perturbations are
observable at the largest scales today. We have provided only
homogeneous solutions. To apply our results to structure formation and
CMB data, the next step will be to begin an analysis of linear density
perturbations in such a Universe.

The model developed here can be extended to cases of less symmetric
evolution. Some examples are models with an aligned B-field plus
cosmic strings neither aligned parallel nor perpendicular to this
field, or models with various kinds of misaligned cosmic
defects. Another possibility with more physical motivation is a model
with B-fields and $\Lambda$ where there are still a few misaligned
magnetic domains within the horizon at the end of inflation. While
this model is more complicated, it may also be more realistic as the
initial domain size is more in line with the models of cosmological
phase transitions
\cite{Savvidy:1977as,Matinian:1976mp,Vachaspati:nm,Enqvist:1994rm,Berera:1998hv}.
Finally, while there could be other combinations of cosmological
constant, magnetic fields, defects, and geometry that arise in brane
world physics and produce asymmetric expansion, we believe it is
important to have considered the simplest well motivated case first.

\section*{Acknowledgments}

We thank Gary Hinshaw for stimulating discussions of the WMAP
data. The work of AB was supported by PPARC, and the work of RVB and
TWK was supported by U.S. DoE grant number DE-FG05-85ER40226.


\begin{thebibliography}{99}
\bibitem{ecc} To simplify formulas, we write results in terms of the
ratio of scale factors $e^a/e^b$, which can be easily expressed in
terms of the true eccentricity.
\end{thebibliography}
\end{document}